# Entanglement Teleportation by Qutrits

Arti Chamoli and C. M. Bhandari
Indian Institute of Information Technology, Allahabad, Deoghat, Jhalwa, Allahabad-211011, India.
Email: achamoli@iiita.ac.in, cmbhandari@yahoo.com

Abstract
Quantum entanglement, like other resources, is now considered to be a resource which can be produced, concentrated if required, transported and consumed. After its inception [1] in 1933, various schemes of quantum state teleportation have been proposed using different types of channels. Not restricting to qubit based systems, qutrit states and channels have also been of considerable interest. In the present paper we investigate the teleportation of an unknown single qutrit state as well as two qutrit state through a three qutrit quantum channel along with the required operations to recover the state. This is further generalized to the case of teleportation of n-qutrit system.

Quantum entanglement being a crucial feature of quantum mechanics is a resource to quantum information processes like quantum teleportation [1], quantum cryptography [2], quantum computation [3], etc. Teleportation in regard to quantum computation has been defined in [4] as the disembodied transport of the quantum state of a system and its correlations across space to another system, where system refers to any single or collective particles of matter or energy such as protons, neutrons, electrons, photons, atoms, ions, etc. Much like other resources quantum entanglement is being considered a resource which can be created, swapped, concentrated, transported and consumed. Quantum teleportation, after its inception Bennett et al. [1], has been extensively studied by several researchers [5-7] incorporating various entangled states as quantum channels. Use of channels with multi qubit entanglement, and channels using non-maximally entangled states have also been examined []. In addition to teleporting unknown qubits and unknown multi-particle states [8-12], quantum protocols to teleport states of n-dimensional systems [13, 14] have been formulated. Recently, teleportation of two unknown entangled qutrits [15] through a single two-qutrit maximally entangled pair was explicitly shown which was further generalized to arbitrary quNits (n-dimensional system).

In this paper, we study the teleportation of an unknown single qutrit state as well as an unknown two qutrit system through a three level GHZ [16] quantum channel along with classical communication. We start with quantum teleportation of an unknown single qutrit state via a three dimensional GHZ state. Suppose Alice wants to teleport an unknown qutrit to Chalie, she does it with the assistance of a third person, Bob. Alice, Bob and Charlie a priory share a three qutrit maximally entangled state, say, in the form

$$|\phi\rangle_{123} = \frac{1}{\sqrt{3}}(|000\rangle + |111\rangle + |222\rangle)_{123}$$

of which qutrits 1, 2, and 3 belong to Alice, Bob and Charlie respectively.

The unknown state that Alice wants to teleport to Bob is of the form,

$$|\psi\rangle_U = (\alpha|0\rangle + \beta|1\rangle + \gamma|2\rangle)_U$$

(2)

Alice starts by combining her qutrit of the shared GHZ state with the unknown qutrit and performs a joint measurement on the two. The complete basis $\{|\Phi_{mn}\rangle_{U1}\}$ with $m,n \in \{0,1,2\}$ for joint measurement includes nine maximally entangled states in the form:

$$|\Phi_{00}\rangle_{U1} = \frac{1}{\sqrt{3}}(|20\rangle + |11\rangle + |02\rangle)_{13}$$

$$|\Phi_{01}\rangle_{U1} = \frac{1}{\sqrt{3}}(|10\rangle + |01\rangle + |22\rangle)_{13}$$

$$|\Phi_{02}\rangle_{U1} = \frac{1}{\sqrt{3}}(|00\rangle + |21\rangle + |12\rangle)_{13}$$

$$|\Phi_{10}\rangle_{U1} = \frac{1}{\sqrt{3}}\left(|20\rangle + e^{\frac{2\pi i}{3}}|11\rangle + e^{-\frac{2\pi i}{3}}|02\rangle\right)_{13}$$

$$|\Phi_{11}\rangle_{U1} = \frac{1}{\sqrt{3}}\left(|10\rangle + e^{\frac{2\pi i}{3}}|01\rangle + e^{-\frac{2\pi i}{3}}|22\rangle\right)_{13}$$

$$|\Phi_{12}\rangle_{U1} = \frac{1}{\sqrt{3}}\left(|00\rangle + e^{\frac{2\pi i}{3}}|21\rangle + e^{-\frac{2\pi i}{3}}|12\rangle\right)_{13}$$

$$|\Phi_{20}\rangle_{U1} = \frac{1}{\sqrt{3}}\left(|20\rangle + e^{-\frac{2\pi i}{3}}|11\rangle + e^{\frac{2\pi i}{3}}|02\rangle\right)_{13}$$

$$|\Phi_{21}\rangle_{U1} = \frac{1}{\sqrt{3}}\left(|10\rangle + e^{-\frac{2\pi i}{3}}|01\rangle + e^{\frac{2\pi i}{3}}|22\rangle\right)_{13}$$

$$|\Phi_{22}\rangle_{U1} = \frac{1}{\sqrt{3}}\left(|00\rangle + e^{-\frac{2\pi i}{3}}|21\rangle + e^{\frac{2\pi i}{3}}|12\rangle\right)_{13}$$

(3)

The effect of joint measurement on particles U and 1 in terms of $|\Phi_{mn}\rangle_{U1}$ can be expressed as

$$\frac{1}{3}[\ |\Phi_{00}\rangle_{U1}(\gamma|00\rangle+\beta|11\rangle+\alpha|22\rangle)_{23}+|\Phi_{01}\rangle_{U1}(\beta|00\rangle+\alpha|11\rangle+\gamma|22\rangle)_{23}+$$

$$|\Phi_{02}\rangle_{U1}(\alpha|00\rangle+\gamma|11\rangle+\beta|22\rangle)_{23}+|\Phi_{10}\rangle_{U1}\left(\gamma|00\rangle+e^{\frac{2\pi i}{3}}\beta|11\rangle+e^{-\frac{2\pi i}{3}}\alpha|22\rangle\right)_{23}+$$

$$|\Phi_{11}\rangle_{U1}\left(\beta|00\rangle+e^{\frac{2\pi i}{3}}\alpha|11\rangle+e^{-\frac{2\pi i}{3}}\gamma|22\rangle\right)_{23}+|\Phi_{12}\rangle_{U1}\left(\alpha|00\rangle+e^{\frac{2\pi i}{3}}\gamma|11\rangle+e^{-\frac{2\pi i}{3}}\beta|22\rangle\right)_{23}+$$

$$|\Phi_{20}\rangle_{U1}\left(\gamma|00\rangle+e^{-\frac{2\pi i}{3}}\beta|11\rangle+e^{\frac{2\pi i}{3}}\alpha|22\rangle\right)_{23}+|\Phi_{21}\rangle_{U1}\left(\beta|00\rangle+e^{-\frac{2\pi i}{3}}\alpha|11\rangle+e^{\frac{2\pi i}{3}}\gamma|22\rangle\right)_{23}+$$

$$|\Phi_{22}\rangle_{U1}\left(\alpha|00\rangle+e^{-\frac{2\pi i}{3}}\gamma|11\rangle+e^{\frac{2\pi i}{3}}\beta|22\rangle\right)_{23}\ ]$$

(4)

The joint measurement on particles U and 1 entangles the two without disturbing the entanglement of particles 2 and 3. Now Bob makes the measurement of his particle in the rotated basis $\left[|\tilde{0}\rangle,|\tilde{1}\rangle,|\tilde{2}\rangle\right]$. The basis can be written as:

$$|\tilde{0}\rangle=\frac{|0\rangle+|1\rangle+|2\rangle}{\sqrt{3}}$$

$$|\tilde{1}\rangle=\frac{|0\rangle+e^{\frac{2\pi i}{3}}|1\rangle+e^{-\frac{2\pi i}{3}}|2\rangle}{\sqrt{3}}$$

$$|\tilde{2}\rangle=\frac{|0\rangle+e^{-\frac{2\pi i}{3}}|1\rangle+e^{\frac{2\pi i}{3}}|2\rangle}{\sqrt{3}}$$

(5)

The combined state of the system after Bob's operation in the rotated basis becomes

$$\frac{1}{3\sqrt{3}}[\ |\Psi_0^0\rangle_3\{|\tilde{0}\rangle_2|\Phi_{00}\rangle_{U1}+|\tilde{1}\rangle_2|\Phi_{10}\rangle_{U1}+|\tilde{2}\rangle_2|\Phi_{20}\rangle_{U1}\}+|\Psi_0^1\rangle_3\{|\tilde{0}\rangle_2|\Phi_{20}\rangle_{U1}+|\tilde{1}\rangle_2|\Phi_{00}\rangle_{U1}+|\tilde{2}\rangle_2|\Phi_{10}\rangle_{U1}\}$$

$$+|\Psi_0^2\rangle_3\{|\tilde{0}\rangle_2|\Phi_{10}\rangle_{U1}+|\tilde{1}\rangle_2|\Phi_{20}\rangle_{U1}+|\tilde{2}\rangle_2|\Phi_{00}\rangle_{U1}\}+|\Psi_1^0\rangle_3\{|\tilde{0}\rangle_2|\Phi_{01}\rangle_{U1}+|\tilde{1}\rangle_2|\Phi_{11}\rangle_{U1}+|\tilde{2}\rangle_2|\Phi_{21}\rangle_{U1}\}$$

$$+|\Psi_1^1\rangle_3\{|\tilde{0}\rangle_2|\Phi_{21}\rangle_{U1}+|\tilde{1}\rangle_2|\Phi_{01}\rangle_{U1}+|\tilde{2}\rangle_2|\Phi_{11}\rangle_{U1}\}+|\Psi_1^2\rangle_3\{|\tilde{0}\rangle_2|\Phi_{11}\rangle_{U1}+|\tilde{1}\rangle_2|\Phi_{21}\rangle_{U1}+|\tilde{2}\rangle_2|\Phi_{01}\rangle_{U1}\}$$

$$+|\Psi_2^0\rangle_3\{|\tilde{0}\rangle_2|\Phi_{02}\rangle_{U1}+|\tilde{1}\rangle_2|\Phi_{12}\rangle_{U1}+|\tilde{2}\rangle_2|\Phi_{22}\rangle_{U1}\}+|\Psi_2^1\rangle_3\{|\tilde{0}\rangle_2|\Phi_{22}\rangle_{U1}+|\tilde{1}\rangle_2|\Phi_{02}\rangle_{U1}+|\tilde{2}\rangle_2|\Phi_{12}\rangle_{U1}\}$$

$$+|\Psi_2^2\rangle_3\{|\tilde{0}\rangle_2|\Phi_{12}\rangle_{U1}+|\tilde{1}\rangle_2|\Phi_{22}\rangle_{U1}+|\tilde{2}\rangle_2|\Phi_{02}\rangle_{U1}\}\ ]$$

(6)

where

$$|\Psi_0^0\rangle_3 = \left(\gamma|0\rangle_3 + \beta|1\rangle_3 + \alpha|2\rangle_3\right)$$

$$|\Psi_0^1\rangle_3 = \left(\gamma|0\rangle_3 + e^{-\frac{2\pi i}{3}}\beta|1\rangle_3 + e^{\frac{2\pi i}{3}}\alpha|2\rangle_3\right)$$

$$|\Psi_0^2\rangle_3 = \left(\gamma|0\rangle_3 + e^{\frac{2\pi i}{3}}\beta|1\rangle_3 + e^{-\frac{2\pi i}{3}}\alpha|2\rangle_3\right)$$

$$|\Psi_1^0\rangle_3 = \left(\beta|0\rangle_3 + \alpha|1\rangle_3 + \gamma|2\rangle_3\right)$$

$$|\Psi_1^1\rangle_3 = \left(\beta|0\rangle_3 + e^{-\frac{2\pi i}{3}}\alpha|1\rangle_3 + e^{\frac{2\pi i}{3}}\gamma|2\rangle_3\right)$$

$$|\Psi_1^2\rangle_3 = \left(\beta|0\rangle_3 + e^{\frac{2\pi i}{3}}\alpha|1\rangle_3 + e^{-\frac{2\pi i}{3}}\gamma|2\rangle_3\right)$$

$$|\Psi_2^0\rangle_3 = \left(\alpha|0\rangle_3 + \gamma|1\rangle_3 + \beta|2\rangle_3\right)$$

$$|\Psi_2^1\rangle_3 = \left(\alpha|0\rangle_3 + e^{-\frac{2\pi i}{3}}\gamma|1\rangle_3 + e^{\frac{2\pi i}{3}}\beta|2\rangle_3\right)$$

$$|\Psi_2^2\rangle_3 = \left(\alpha|0\rangle_3 + e^{\frac{2\pi i}{3}}\gamma|1\rangle_3 + e^{-\frac{2\pi i}{3}}\beta|2\rangle_3\right)$$

(7)

Alice's joint measurement on particles U and 1 will result into entanglement of particles *U* and *1* in any one of the nine maximally entangled forms as expressed in eq. (3). Alice's measurement followed by Bob's measurement in the rotated basis will leave the state of all four particles in one of the nine forms given by eq. (4). It can be seen in eq.(6) that Bob's measurement leads to the association of unknown coefficients $(\alpha, \beta, \gamma)$ to Charlie's particle. Expression in eq. (6) shows possible outcomes, the system can transform to after Alice and Bob's measurement of their respective particles. There are 27 different outcomes with equal probabilities. The result of Bob's measurement of particle *2* can be specified by *l*, where *l* = 0 (1, 2) corresponds to finding particle 2 in state $|\tilde{0}\rangle_2 (|\tilde{1}\rangle_2, |\tilde{2}\rangle_2)$. The result of Alice's joint measurement on particles U and 1 can be specified by $|\Phi_{mn}\rangle_{U1}$, where {*m, n*}={0, 0} ({0, 1}, {0, 2}…..) correspond to finding the state of particles U and 1 as $|\Phi_{00}\rangle_{U1} (|\Phi_{01}\rangle_{U1}, |\Phi_{02}\rangle_{U1}, ......)$. Alice then classically announces which of the nine maximally entangled states she ended up projecting her particles. This is followed by Bob's announcement of his measurement in the rotated basis. Depending on Alice's measurement outcome and Bob's as well, Charlie will use a suitable operator to convert his qutrit into the state (2). Table1 shows the possible outcomes of Alice's and Bob's measurement and the corresponding unitary operator, $U_{lmn}$, Charlie needs to apply in order to successfully construct the teleported state.

Table1. List of Charlie's unitary operations based on Alice's and Bob's measurement outcomes, $l$ and $|\Phi_{mn}\rangle_{23}$ respectively.

| $l$ | $|\Phi_{mn}\rangle_{U1}$ | $U_{lmn}$ |
|---|---|---|
| 0(1,2) | $|\Phi_{00}\rangle_{U1} (|\Phi_{10}\rangle_{U1}, |\Phi_{20}\rangle_{U1})$ | $|2\rangle_3\langle 0| + |1\rangle_3\langle 1| + |0\rangle_3\langle 2|$ |
| 0(1,2) | $|\Phi_{20}\rangle_{U1} (|\Phi_{00}\rangle_{U1}, |\Phi_{10}\rangle_{U1})$ | $|2\rangle_3\langle 0| + e^{\frac{2\pi i}{3}}|1\rangle_3\langle 1| + e^{\frac{-2\pi i}{3}}|0\rangle_3\langle 2|$ |
| 0(1,2) | $|\Phi_{10}\rangle_{U1} (|\Phi_{20}\rangle_{U1}, |\Phi_{00}\rangle_{U1})$ | $|2\rangle_3\langle 0| + e^{\frac{-2\pi i}{3}}|1\rangle_3\langle 1| + e^{\frac{2\pi i}{3}}|0\rangle_3\langle 2|$ |
| 0(1,2) | $|\Phi_{01}\rangle_{U1} (|\Phi_{11}\rangle_{U1}, |\Phi_{21}\rangle_{U1})$ | $|1\rangle_3\langle 0| + |0\rangle_3\langle 1| + |2\rangle_3\langle 2|$ |
| 0(1,2) | $|\Phi_{21}\rangle_{U1} (|\Phi_{01}\rangle_{U1}, |\Phi_{11}\rangle_{U1})$ | $|1\rangle_3\langle 0| + e^{\frac{2\pi i}{3}}|0\rangle_3\langle 1| + e^{\frac{-2\pi i}{3}}|2\rangle_3\langle 2|$ |
| 0(1,2) | $|\Phi_{11}\rangle_{U1} (|\Phi_{21}\rangle_{U1}, |\Phi_{01}\rangle_{U1})$ | $|1\rangle_3\langle 0| + e^{\frac{-2\pi i}{3}}|0\rangle_3\langle 1| + e^{\frac{2\pi i}{3}}|2\rangle_3\langle 2|$ |
| 0(1,2) | $|\Phi_{02}\rangle_{U1} (|\Phi_{12}\rangle_{U1}, |\Phi_{22}\rangle_{U1})$ | $|0\rangle_3\langle 0| + |2\rangle_3\langle 1| + |1\rangle_3\langle 2|$ |
| 0(1,2) | $|\Phi_{22}\rangle_{U1} (|\Phi_{02}\rangle_{U1}, |\Phi_{12}\rangle_{U1})$ | $|0\rangle_3\langle 0| + e^{\frac{2\pi i}{3}}|2\rangle_3\langle 1| + e^{\frac{-2\pi i}{3}}|1\rangle_3\langle 2|$ |
| 0(1,2) | $|\Phi_{12}\rangle_{U1} (|\Phi_{22}\rangle_{U1}, |\Phi_{02}\rangle_{U1})$ | $|0\rangle_3\langle 0| + e^{\frac{-2\pi i}{3}}|2\rangle_3\langle 1| + e^{\frac{2\pi i}{3}}|1\rangle_3\langle 2|$ |

**Teleporting an unknown two qutrit system**

Suppose Alice wants to teleport an unknown two qutrit system in entangled state
$$|\psi\rangle_{12} = (\alpha|00\rangle + \beta|11\rangle + \gamma|22\rangle)_{12}$$
(8)

where $|\alpha|^2 + |\beta|^2 + |\gamma|^2 = 1$. Alice and Bob initially share a maximally entangled three qutrit system in the form
$$|\phi\rangle_{345} = \frac{1}{\sqrt{3}}(|000\rangle + |111\rangle + |222\rangle)_{345}$$
(9)

of which qutrit 3 is with Alice and qutrits 4 and 5 are with Bob. Alice combines the unknown qutrit system with her particle of the shared system. The combined state is in the form

$$|\Psi\rangle_{12345} = \frac{\alpha}{\sqrt{3}}\left(|00000\rangle+|00111\rangle+|00222\rangle\right)_{12345} + \frac{\beta}{\sqrt{3}}\left(|11000\rangle+|11111\rangle+|11222\rangle\right)_{12345}$$
$$+ \frac{\gamma}{\sqrt{3}}\left(|22000\rangle+|22111\rangle+|22222\rangle\right)_{12345}$$

(10)

Alice performs joint measurement on her particles 2 and 3 in the basis $|\Phi_{mn}\rangle$ as given by (3). The total state of the combined system in terms of $|\Phi_{mn}\rangle_{23}$ system after performing joint measurement becomes

$$\frac{1}{3}\Big[|\Phi_{00}\rangle_{23}\left(\gamma|200\rangle+\beta|111\rangle+\alpha|022\rangle\right)_{145} + |\Phi_{01}\rangle_{23}\left(\beta|100\rangle+\alpha|011\rangle+\gamma|222\rangle\right)_{145}$$
$$|\Phi_{02}\rangle_{23}\left(\alpha|000\rangle+\gamma|211\rangle+\beta|122\rangle\right)_{145} + |\Phi_{10}\rangle_{23}\left(\gamma|200\rangle+e^{\frac{2\pi i}{3}}\beta|211\rangle+e^{\frac{-2\pi i}{3}}\alpha|022\rangle\right)_{145}$$
$$|\Phi_{11}\rangle_{23}\left(\beta|100\rangle+e^{\frac{2\pi i}{3}}\alpha|011\rangle+e^{\frac{-2\pi i}{3}}\gamma|222\rangle\right)_{145} + |\Phi_{12}\rangle_{23}\left(\alpha|000\rangle+e^{\frac{2\pi i}{3}}\gamma|211\rangle+e^{\frac{-2\pi i}{3}}\beta|122\rangle\right)_{145}$$
$$|\Phi_{20}\rangle_{23}\left(\gamma|200\rangle+e^{\frac{-2\pi i}{3}}\beta|111\rangle+e^{\frac{2\pi i}{3}}\alpha|022\rangle\right)_{145} + |\Phi_{21}\rangle_{23}\left(\beta|100\rangle+e^{\frac{-2\pi i}{3}}\alpha|011\rangle+e^{\frac{2\pi i}{3}}\gamma|222\rangle\right)_{145}$$
$$+|\Phi_{22}\rangle_{23}\left(\alpha|000\rangle+e^{\frac{-2\pi i}{3}}\gamma|211\rangle+e^{\frac{2\pi i}{3}}\beta|122\rangle\right)_{145}\Big]$$

(10)

The effect of joint measurement on the particles 2 and 3 makes them entangled. Consequently, particles 1, 4 and 5 become entangled. Now Alice makes a second measurement of particle 1 in the rotated basis $\left[|\tilde{0}\rangle,|\tilde{1}\rangle,|\tilde{2}\rangle\right]$. The detailed expression for the same has been given in the previous section. The effect of Alice's second measurement on the combined state can be expressed as

$$\frac{1}{3\sqrt{3}}\Big[|\Psi_0^0\rangle_{45}\left\{|\tilde{0}\rangle_1|\Phi_{00}\rangle_{23}+|\tilde{1}\rangle_1 e^{\frac{2\pi i}{3}}|\Phi_{20}\rangle_{23}+|\tilde{2}\rangle_1 e^{\frac{-2\pi i}{3}}|\Phi_{10}\rangle_{23}\right\} + |\Psi_0^1\rangle_{45}\left\{|\tilde{0}\rangle_1|\Phi_{10}\rangle_{23}+|\tilde{1}\rangle_1 e^{\frac{2\pi i}{3}}|\Phi_{00}\rangle_{23}+|\tilde{2}\rangle_1 e^{\frac{-2\pi i}{3}}|\Phi_{20}\rangle_{23}\right\}$$
$$+|\Psi_0^2\rangle_{45}\left\{|\tilde{0}\rangle_1|\Phi_{20}\rangle_{23}+|\tilde{1}\rangle_1 e^{\frac{2\pi i}{3}}|\Phi_{10}\rangle_{23}+|\tilde{2}\rangle_1 e^{\frac{-2\pi i}{3}}|\Phi_{00}\rangle_{23}\right\} + |\Psi_1^0\rangle_{45}\left\{|\tilde{0}\rangle_1|\Phi_{01}\rangle_{23}+|\tilde{1}\rangle_1 e^{\frac{-2\pi i}{3}}|\Phi_{21}\rangle_{23}+|\tilde{2}\rangle_1 e^{\frac{2\pi i}{3}}|\Phi_{11}\rangle_{23}\right\}$$
$$+|\Psi_1^1\rangle_{45}\left\{|\tilde{0}\rangle_1|\Phi_{11}\rangle_{23}+|\tilde{1}\rangle_1 e^{\frac{-2\pi i}{3}}|\Phi_{01}\rangle_{23}+|\tilde{2}\rangle_1 e^{\frac{2\pi i}{3}}|\Phi_{21}\rangle_{23}\right\} + |\Psi_1^2\rangle_{45}\left\{|\tilde{0}\rangle_1|\Phi_{21}\rangle_{23}+|\tilde{1}\rangle_1 e^{\frac{-2\pi i}{3}}|\Phi_{11}\rangle_{23}+|\tilde{2}\rangle_1 e^{\frac{2\pi i}{3}}|\Phi_{01}\rangle_{23}\right\}$$
$$+|\Psi_2^0\rangle_{45}\left\{|\tilde{0}\rangle_1|\Phi_{02}\rangle_{23}+|\tilde{1}\rangle_1|\Phi_{22}\rangle_{23}+|\tilde{2}\rangle_1|\Phi_{12}\rangle_{23}\right\} + |\Psi_2^1\rangle_{45}\left\{|\tilde{0}\rangle_1|\Phi_{12}\rangle_{23}+|\tilde{1}\rangle_1|\Phi_{02}\rangle_{23}+|\tilde{2}\rangle_1|\Phi_{22}\rangle_{23}\right\}$$
$$+|\Psi_2^2\rangle_{45}\left\{|\tilde{0}\rangle_1|\Phi_{22}\rangle_{23}+|\tilde{1}\rangle_1|\Phi_{12}\rangle_{23}+|\tilde{2}\rangle_1|\Phi_{02}\rangle_{23}\right\}\Big]$$

(11)

where

$$|\Psi_0^0\rangle_{45} = (\gamma|00\rangle_{45} + \beta|11\rangle_{45} + \alpha|22\rangle_{45})$$

$$|\Psi_0^1\rangle_{45} = \left(\gamma|00\rangle_{45} + e^{\frac{2\pi i}{3}}\beta|11\rangle_3 + e^{\frac{-2\pi i}{3}}\alpha|22\rangle_3\right)$$

$$|\Psi_0^2\rangle_{45} = \left(\gamma|00\rangle_{45} + e^{\frac{-2\pi i}{3}}\beta|11\rangle_{45} + e^{\frac{2\pi i}{3}}\alpha|22\rangle_{45}\right)$$

$$|\Psi_1^0\rangle_{45} = (\beta|00\rangle_{45} + \alpha|11\rangle_{45} + \gamma|22\rangle_{45})$$

$$|\Psi_1^1\rangle_{45} = \left(\beta|00\rangle_{45} + e^{\frac{2\pi i}{3}}\alpha|11\rangle_{45} + e^{\frac{-2\pi i}{3}}\gamma|22\rangle_{45}\right)$$

$$|\Psi_1^2\rangle_{45} = \left(\beta|00\rangle_{45} + e^{\frac{-2\pi i}{3}}\alpha|11\rangle_{45} + e^{\frac{2\pi i}{3}}\gamma|22\rangle_{45}\right)$$

$$|\Psi_2^0\rangle_{45} = (\alpha|00\rangle_{45} + \gamma|11\rangle_{45} + \beta|22\rangle_{45})$$

$$|\Psi_2^1\rangle_{45} = \left(\alpha|00\rangle_{45} + e^{\frac{2\pi i}{3}}\gamma|11\rangle_{45} + e^{\frac{-2\pi i}{3}}\beta|22\rangle_{45}\right)$$

$$|\Psi_2^2\rangle_{45} = \left(\alpha|00\rangle_{45} + e^{\frac{-2\pi i}{3}}\gamma|11\rangle_{45} + e^{\frac{2\pi i}{3}}\beta|22\rangle_{45}\right)$$

(12)

Similar to the previous case, Alice's measurement has 27 possible outcomes with equal probabilities. The measurement result of particle 1 can be expressed as $l = 0$ (1, 2) corresponds to finding particle 1 in state $|\tilde{0}\rangle_1$ ($|\tilde{1}\rangle_1, |\tilde{2}\rangle_1$) and that of joint measurement on particles 2 and 3 as $|\Phi_{mn}\rangle_{23} = |\Phi_{00}\rangle_{23}(|\Phi_{01}\rangle_{23}, |\Phi_{02}\rangle_{23}, \ldots)$. Alice will classically inform Bob about her result. Depending on Alice's measurement outcomes Bob will apply specific unitary operator, $U_{lmn}$, as illustrated in table2.

Table2. List of Bob's unitary operations based on Alice's measurement outcomes $l$ and $|\Phi_{mn}\rangle_{23}$.

| $l$ | $|\Phi_{mn}\rangle_{23}$ | $U_{lmn}$ |
|---|---|---|
| 0(1,2) | $|\Phi_{00}\rangle_{23}(|\Phi_{20}\rangle_{23}, |\Phi_{10}\rangle_{23})$ | $|22\rangle_{45}\langle 00| + |11\rangle_{45}\langle 11| + |00\rangle_{45}\langle 22|$ |
| 0(1,2) | $|\Phi_{10}\rangle_{23}(|\Phi_{00}\rangle_{23}, |\Phi_{20}\rangle_{23})$ | $|22\rangle_{45}\langle 00| + e^{\frac{-2\pi i}{3}}|11\rangle_{45}\langle 11| + e^{\frac{2\pi i}{3}}|00\rangle_{45}\langle 22|$ |

| | | |
|---|---|---|
| 0(1,2) | $\|\Phi_{20}\rangle_{23}(\|\Phi_{10}\rangle_{23},\|\Phi_{00}\rangle_{23})$ | $\|22\rangle_{45}\langle 00\| + e^{\frac{2\pi i}{3}}\|11\rangle_{45}\langle 11\| + e^{\frac{-2\pi i}{3}}\|00\rangle_{45}\langle 22\|$ |
| 0(1,2) | $\|\Phi_{01}\rangle_{23}(\|\Phi_{21}\rangle_{23},\|\Phi_{11}\rangle_{23})$ | $\|11\rangle_{45}\langle 00\| + \|00\rangle_{45}\langle 11\| + \|22\rangle_{45}\langle 22\|$ |
| 0(1,2) | $\|\Phi_{11}\rangle_{23}(\|\Phi_{01}\rangle_{23},\|\Phi_{21}\rangle_{23})$ | $\|11\rangle_{45}\langle 00\| + e^{\frac{-2\pi i}{3}}\|00\rangle_{45}\langle 11\| + e^{\frac{2\pi i}{3}}\|22\rangle_{45}\langle 22\|$ |
| 0(1,2) | $\|\Phi_{21}\rangle_{23}(\|\Phi_{11}\rangle_{23},\|\Phi_{01}\rangle_{23})$ | $\|11\rangle_{45}\langle 00\| + e^{\frac{2\pi i}{3}}\|00\rangle_{45}\langle 11\| + e^{\frac{-2\pi i}{3}}\|22\rangle_{45}\langle 22\|$ |
| 0(1,2) | $\|\Phi_{02}\rangle_{23}(\|\Phi_{22}\rangle_{23},\|\Phi_{12}\rangle_{23})$ | $\|00\rangle_{45}\langle 00\| + \|22\rangle_{45}\langle 11\| + \|11\rangle_{45}\langle 22\|$ |
| 0(1,2) | $\|\Phi_{12}\rangle_{23}(\|\Phi_{02}\rangle_{23},\|\Phi_{22}\rangle_{23})$ | $\|00\rangle_{45}\langle 00\| + e^{\frac{-2\pi i}{3}}\|22\rangle_{45}\langle 11\| + e^{\frac{2\pi i}{3}}\|11\rangle_{45}\langle 22\|$ |
| 0(1,2) | $\|\Phi_{22}\rangle_{23}(\|\Phi_{12}\rangle_{23},\|\Phi_{02}\rangle_{23})$ | $\|00\rangle_{45}\langle 00\| + e^{\frac{2\pi i}{3}}\|22\rangle_{45}\langle 11\| + e^{\frac{-2\pi i}{3}}\|11\rangle_{45}\langle 22\|$ |

We can now generalize the above scheme to the teleportation of unknown n-qutrit system. Any unknown *n*- qutrit system can be teleported through a quantum channel of maximally entangled (*n*+1)-qutrits. Let the unknown state Alice wishes to teleport to Bob is given by

$$|\Psi\rangle_{1,2,\ldots n} = \alpha|0_1 0_2 \ldots 0_n\rangle + \beta|1_1 1_2 \ldots 1_n\rangle + \gamma|2_1 2_2 \ldots 2_n\rangle$$

(13)

where $|\alpha|^2 + |\beta|^2 + |\gamma|^2 = 1$. Alice and Bob a priory share a maximally entangled state of (*n*+1) qutrits which serves as the quantum channel for successful teleportation. The shared maximally entangled state is such that its only one particle is kept by Alice and rest *n* particles are with Bob. The maximally entangled state is in the form, say,

$$|\Phi\rangle_{1',2',\ldots(n+1)'} = \frac{1}{\sqrt{3}}\left(|0_{1'} 0_{2'} \ldots 0_{(n+1)'}\rangle + |1_{1'} 1_{2'} \ldots 1_{(n+1)'}\rangle + |2_{1'} 2_{2'} \ldots 2_{(n+1)'}\rangle\right)$$

(14)

(*n*+1)-qutrits which constitute the maximally entangled state have been represented with dashed subscripts as $(0_{1'} 0_{2'} \ldots 0_{(n+1)'})$ whereas *n*-qutrits of the unknown state to be teleported have been represented as $(0_1 0_2 \ldots 0_n)$.

Suppose Alice keeps particle $1'$ of the maximally entangled state. She combines $1'$ with the state to be teleported. The combined state looks like

$$\frac{\alpha}{\sqrt{3}}\left|0_1 0_2 ...... 0_n 0_{1'} 0_{2'} ...... 0_{(n+1)'}\right\rangle + \frac{\beta}{\sqrt{3}}\left|1_1 1_2 ...... 1_n 1_{1'} 1_{2'} ...... 1_{(n+1)'}\right\rangle + \frac{\gamma}{\sqrt{3}}\left|2_1 2_2 ...... 2_n 2_{1'} 2_{2'} ...... 2_{(n+1)'}\right\rangle$$
(15)

She then performs a joint measurement on particle $1'$ and $n$th particle of the state to be teleported. This projects $(1'-n)$ system to one of the maximally entangled pair as given by (3). Consequently the remaining particles become entangled as well. She measure particles 1 through $(n-1)$ of the state to be teleported in the rotated basis $\left[\left|\tilde{0}\right\rangle, \left|\tilde{1}\right\rangle, \left|\tilde{2}\right\rangle\right]$. Alice publicly then announces the results of her measurements. Alice's joint measurement on $1'$ and $n$th particle results into nine possible states, consequently Bob's particles will be projected to nine orthogonal states. Accordingly, Bob performs one of the following unitary transformations in order to convert his n-qutrits into state (13).

$$\left|2_{2'} 2_{3'}......2_{(n+1)'}\right\rangle\left\langle 0_{2'} 0_{3'}......0_{(n+1)'}\right| + \left|1_{2'} 1_{3'}......1_{(n+1)'}\right\rangle\left\langle 1_{2'} 1_{3'}......1_{(n+1)'}\right| + \left|0_{2'} 0_{3'}......0_{(n+1)'}\right\rangle\left\langle 2_{2'} 2_{3'}......2_{(n+1)'}\right|$$

$$\left|2_{2'} 2_{3'}......2_{(n+1)'}\right\rangle\left\langle 0_{2'} 0_{3'}......0_{(n+1)'}\right| + e^{\frac{-2\pi i}{3}}\left|1_{2'} 1_{3'}......1_{(n+1)'}\right\rangle\left\langle 1_{2'} 1_{3'}......1_{(n+1)'}\right| + e^{\frac{2\pi i}{3}}\left|0_{2'} 0_{3'}......0_{(n+1)'}\right\rangle\left\langle 2_{2'} 2_{3'}......2_{(n+1)'}\right|$$

$$\left|2_{2'} 2_{3'}......2_{(n+1)'}\right\rangle\left\langle 0_{2'} 0_{3'}......0_{(n+1)'}\right| + e^{\frac{2\pi i}{3}}\left|1_{2'} 1_{3'}......1_{(n+1)'}\right\rangle\left\langle 1_{2'} 1_{3'}......1_{(n+1)'}\right| + e^{\frac{-2\pi i}{3}}\left|0_{2'} 0_{3'}......0_{(n+1)'}\right\rangle\left\langle 2_{2'} 2_{3'}......2_{(n+1)'}\right|$$

$$\left|1_{2'} 1_{3'}......1_{(n+1)'}\right\rangle\left\langle 0_{2'} 0_{3'}......0_{(n+1)'}\right| + \left|0_{2'} 0_{3'}......0_{(n+1)'}\right\rangle\left\langle 1_{2'} 1_{3'}......1_{(n+1)'}\right| + \left|2_{2'} 2_{3'}......2_{(n+1)'}\right\rangle\left\langle 2_{2'} 2_{3'}......2_{(n+1)'}\right|$$

$$\left|1_{2'} 1_{3'}......1_{(n+1)'}\right\rangle\left\langle 0_{2'} 0_{3'}......0_{(n+1)'}\right| + e^{\frac{-2\pi i}{3}}\left|0_{2'} 0_{3'}......0_{(n+1)'}\right\rangle\left\langle 1_{2'} 1_{3'}......1_{(n+1)'}\right| + e^{\frac{2\pi i}{3}}\left|2_{2'} 2_{3'}......2_{(n+1)'}\right\rangle\left\langle 2_{2'} 2_{3'}......2_{(n+1)'}\right|$$

$$\left|1_{2'} 1_{3'}......1_{(n+1)'}\right\rangle\left\langle 0_{2'} 0_{3'}......0_{(n+1)'}\right| + e^{\frac{2\pi i}{3}}\left|0_{2'} 0_{3'}......0_{(n+1)'}\right\rangle\left\langle 1_{2'} 1_{3'}......1_{(n+1)'}\right| + e^{\frac{-2\pi i}{3}}\left|2_{2'} 2_{3'}......2_{(n+1)'}\right\rangle\left\langle 2_{2'} 2_{3'}......2_{(n+1)'}\right|$$

$$\left|0_{2'} 0_{3'}......0_{(n+1)'}\right\rangle\left\langle 0_{2'} 0_{3'}......0_{(n+1)'}\right| + \left|22\right\rangle\left\langle 1_{2'} 1_{3'}......1_{(n+1)'}\right| + \left|1_{2'} 1_{3'}......1_{(n+1)'}\right\rangle\left\langle 2_{2'} 2_{3'}......2_{(n+1)'}\right|$$

$$\left|0_{2'} 0_{3'}......0_{(n+1)'}\right\rangle\left\langle 0_{2'} 0_{3'}......0_{(n+1)'}\right| + e^{\frac{-2\pi i}{3}}\left|2_{2'} 2_{3'}......2_{(n+1)'}\right\rangle\left\langle 1_{2'} 1_{3'}......1_{(n+1)'}\right| + e^{\frac{2\pi i}{3}}\left|1_{2'} 1_{3'}......1_{(n+1)'}\right\rangle\left\langle 2_{2'} 2_{3'}......2_{(n+1)'}\right|$$

$$\left|0_{2'} 0_{3'}......0_{(n+1)'}\right\rangle\left\langle 0_{2'} 0_{3'}......0_{(n+1)'}\right| + e^{\frac{2\pi i}{3}}\left|2_{2'} 2_{3'}......2_{(n+1)'}\right\rangle\left\langle 1_{2'} 1_{3'}......1_{(n+1)'}\right| + e^{\frac{-2\pi i}{3}}\left|1_{2'} 1_{3'}......1_{(n+1)'}\right\rangle\left\langle 2_{2'} 2_{3'}......2_{(n+1)'}\right|$$
(16)

We have studied the teleportation of unknown one and two qutrit system through a quantum channel of maximally entangled three qutit system. Later we have generalized the same scheme for teleporting unknown $n$-qutrit system using a maximally entangled quantum channel of $(n+1)$-qutrits. Teleportation of any number of qutrits is possible by making a joint measurement in the basis consisting of nine maximally entangled pairs of qutrits (3) followed by a measurement in the rotated basis $\left[\left|\tilde{0}\right\rangle, \left|\tilde{1}\right\rangle, \left|\tilde{2}\right\rangle\right]$. The announcement of measurement outcomes by Alice lets Bob to apply the appropriate unitary operator so as to retrieve the unknown state exactly. Nguyen [15] has shown that Alice teleports two qutrits in an entangled state through a two-qutrit maximally entangled pair shared by her and Bob. In the process, Bob, the receiver, has to prepare an ancilla qutrit in state $|0\rangle$ and apply control change gate [15] on ancilla qutrit and his qutrit of the shared maximally entangled pair. This is done to get ancilla qutrit entangled with the combined state of Alice and his particles. Whereas using a GHZ state as a quantum channel for teleporting two qutrits entangled state, or in general, ($n$+1) maximally entangled pair for teleporting $n$- qutrit state makes teleportation possible without introducing ancilla qutrit and consequent application of control change gate.